\documentstyle[12pt]{article}

\input epsf
\begin{document}
\setcounter{page}{1}
\begin{titlepage}
\hfill Preprint YerPhI-1477(14)-96
\vspace{2cm}

\vspace{2cm}
\begin{center}

{\bf
Triplet Higgs Bosons Production in $e^-e^-$-Collisions}\\
\vspace{5mm}
{\large R.A. Alanakyan}\\
\vspace{5mm}
{\em Theoretical Physics Department,
Yerevan Physics Institute,
Alikhanian Brothers St.2,

 Yerevan 375036, Armenia\\}
 {E-mail: alanak @ lx2.yerphi.am\\}
\end{center}

\vspace{5mm}
\centerline{{\bf{Abstract}}}
In the framework of models with Higgs triplets, doubly and singly 
charged triplet Higgs boson production in the processes
 $e^-e^-\rightarrow\delta ^{--}_{L,R}Z^0$ and  
 $e^-e^-\rightarrow\delta ^{-}_{L}W^-_L$  
are considered.
\vspace{5mm}

{\em keywords:Higgs  12.15.Cc,triplet 14.80.-j,collisions

13.,neutrino 13.15.-f,violation 11.30.Hv \\}

\vfill
\centerline{{\bf{Yerevan Physics Institute}}}
\centerline{{\bf{Yerevan 1998}}}

\end{titlepage}

                 {\bf 1.Introduction}

As known, isospin triplets (with $T=1$  and hypercharge $Y=2$) of Higgs bosons
 provide a natural explanation of the
smallness of the left-handed neutrino masses by See-Saw mechanism 
(see e.g. \cite{GRS}-\cite{G}
and references therein).

 Another consequence of such Higgs triplet introduction is the presence
 of some 
new phenomena with lepton number violation such as neutrinoless $\beta$
-decays, $\mu\rightarrow 3e$ decay, muonium-antimuonium conversion
 mediated by doubly charged Higgs bosons and/or  Majorana
neutrinos (see refs.\cite{G}-\cite{FC} and references
therein).
 
In $e^-e^-$ collisions
 doubly charged Higgs bosons
  may be produced at lower order in resonance \cite{TR},\cite{JG}:
\begin{equation}
\label{A1}
 e^-e^-\rightarrow \delta^{--}_{L,R},
\end{equation}

with subsequent decays   $\delta^{--}_{L,R}\rightarrow l^-l^-$,
$W^-_{L,R}W^-_{L,R}$ ,$W^-_{L}\delta^{-}_{L,}$.

It must be noted however, that the mass of the Higgs bosons is the free
 parameter of the theory and, thus, we don't know  which energies are
necessary for  $\delta^{--}_{L,R}$-bosons production
 in resonance
 whereas  doubly charged Higgs bosons production in association with other
  particles may
soften the above resonanse condition.For instance in ref.
\cite{RA} has been considered \footnote{Previously the process (2)
 has been considered in ref.\cite{TR}
however in this reference
 the third $s$-channel diagramm of Fig.1  which is nesessary for gauge
 invariance of the process (2) has been neglected
 and the cross section of this reference ( formula (3.31))
 differs from  result of ref.\cite{RA} (formula(4)).}
 the process (see Fig.1):

\begin{equation}
\label{A2}
 e^-e^-\rightarrow \delta^{--}_{L,R}\gamma,
\end{equation}
which is allowed if $\sqrt{s}>m_H$.

Here we study
 doubly charged  Higgs bosons production 
in the process:

\begin{equation}
\label{A33}
 e^-e^-\rightarrow \delta^{--}_{L,R}Z^0,
\end{equation}
and the process of  singly charged triplet Higgs boson (which is also a
 member of the left Higgs triplet) production:

\begin{equation}
\label{A4}
 e^-e^-\rightarrow \delta^{-}_{L}W^-_L.
\end{equation}
These processes are described by the three diagrams of Fig.1.
Previously, the process (4)
 has been considered in ref.\cite{H}
however, there
 the third $s$-channel diagramm of Fig.1 with virtual
  $\delta^{--}_L$-boson exchange has not been considered
and the cross section (formula (5.1)) differs from
our result expressed by formulas (9)-(15) below, where this
 $s$-channel $\delta^{--}_L$-boson exchange is taken into account.

It must  be noted that whereas reaction (1) is the lowest order
of the  doubly charged Higgs bosons,
 reaction (4) is the lowest order of singly
charged Higgs bosons production.

 Singly charged triplet Higgs  bosons,
produced in reaction (4)
as  well as doubly charged Higgs  bosons may decay  into leptons
( $\delta^{-}_{L} \rightarrow l^-\nu$)
or into gauge  bosons ( e.g. into  $W^-_{L}Z^0$) or into
gauge bosons and/or more light  Higgs bosons (these decays 
are considered in ref. \cite{G}).

If $Z^0$-bosons produced in reaction (3) decay into neutrino pairs the final
 state of the process (3)
 is the same as in reaction of 
$W^-_LW^-_L$-fusion \cite{T}:
\begin{equation}
\label{A30}
 e^-e^-\rightarrow \delta^{--}_{L}\nu_e\nu_e.
\end{equation}
 The cross section
 of the process (5) via $W^-_LW^-_L$-fusion is of the same order as the
cross section
of the standard $H^0$- boson  production by 
$W^+_LW^-_L$-
fusion considered in
ref.\cite{CD}- \cite{D} multiplied by $(\frac{v_L}{k_L})^2$ because the vertex
$W^+_LW^-_LH^0$ multiplied by $\frac{v_L}{k_L}$ is of the same
order as the vertex  $W_L^- W_L^- \delta^{++}_{L}.$

Far from the threshold the cross section of the
$W^-_LW^-_L$-fusion increases as
:
\begin{equation}
\label{A51}
\sigma=\frac{1}{16\sin^6\theta_W} \frac{ \alpha^3}{m^2_W}
(\frac{v_L}{k_L})^2
(\log(\frac{s}{m^2_H})-2),
\end{equation}
whereas the cross section of the processes (3),(4) decreases as
$s^{-1}$ with growth of $\sqrt{s}$.

In the left-right symmetric model \cite{PS}-\cite{S4},
\cite{M1}-\cite{G},\cite{GVW} with Higgs triplets
\cite{M1}-\cite{G},\cite{GVW}
 (see Appendix), however, the vertex $W_L^-W_L^-\delta^{++}_{L}$
which is responsible for $W_L^-W_L^-$-fusion is suppresed
by the factor of $\frac{v_L}{k_L}$, which must be small
 to preserve the true relation between
$W^-_L$,$Z^0$-bosons masses and Weinberg's angle.Besides, because
$Z^0$-boson is on shell, the process (5)
mediated by the mechanism (3) with subsequent decay of $Z^0$-boson
into neutrino pairs has in fact 
a 2-body final state and that is why it does not decrease
with the growth of $m_H$ as fast as the process (5)
mediated by 
$W^-_L W^-_L $-fusion which has a 3-particle final state.

Thus, at sufficiently small
 $\frac{v_L}{k_L}$ \footnote{In other models with Higgs triplets
 factor $\frac{v_L}{k_L}$ in vertex
 $W_L^-W_L^-\delta_L^{++} $
  may be not small
 (e.g. in the standard model where Higgs sector contains also
 two Higgs triplets with $Y=2;0$, see
 refs.\cite{GM}-\cite{GVW})  and
 the  role of $W^-_LW^-_L$-fusion  in the process
 (5) increases
 in comparison with the  left-right  symmetric model with Higgs
 triplets  case.
 Excluding large  factor $\frac{v_L}{k_L}$ case
all our results  are also applicable to any model
 with left and/or right Higgs triplets.}, sufficiently large
 Yucawa couplings $h_{ee}$ and not
 very high energies the studied
 contribution dominates over $W^-_LW^-_L$-fusion.

For instance, at
 $\frac{v_L}{k_L}=10^{-2}$,
$h_{ee}=10^{-2}$,$\sqrt{s}=1$ TeV and $m_H=100$ GeV the contribution
 of the diagramms of Fig.1 exceeds the
 contribution of $W^-_LW^-_L$-fusion approximately by a factor 100.

Analogously, if $W^-_L$-bosons produced in reaction (4)
 decay into $e^-\nu_e$
we must in general consider jointly the process (4) with subsequent
 decays $W^-_L\rightarrow e^-\nu_e$
and $\delta_L^-$-bosons production via 
vertex 
$W_L^-Z^0\delta_L^+$ in $Z^0W^-_L$ fusion \cite{R},\cite{GMN} ( which is of the
 same order as vertex $W_L^-W_L^-\delta_L^{++} $)
 as parts of the same process:
\begin{equation}
\label{A6}
 e^-e^-\rightarrow \delta^{-}_{L}e^-\nu_e.
\end{equation}
All conclusions concerning reaction (5) are also true in
the case of reaction (7),
because vertex 
$W_L^-Z^0\delta_L^+$ is
of the same order as vertex $W_L^-W_L^-\delta_L^{++} $ also small
in the framework of the left-right symmetric model.
 
{\bf 2.Results}

Using formula (A5) in Appendix A for the
$\delta^{--}_{L,R}$-interaction with electrons we obtain the
following amplitudes
of processes (3),(4):
\begin{equation}
\label{A7}
M=2eh_{ee}a_{L,R}\bar{u}^c(k_1)\left( \frac{\hat{k}_4
\hat{Z}}{t}+\frac{\hat{Z}\hat{k}_4}
{u}+4\frac{(k_4Z)}{s-m_H^2}\right) P_{L,R}u(k_2),
\end{equation}
\begin{equation}
\label{FF}
M=\frac{eh_{ee}}{\sin\theta_W}\bar{u}^c(k_1)\left( \frac{\hat{k}_4
\hat{W}}{t}+\frac{\hat{W}\hat{k}_4}
{u}+4\frac{(k_4W)}{s-m_H^2+im_H \Gamma_H}\right) P_{L}u(k_2).
\end{equation}   
Here we neglect the electron mass and use the following notation:
 $Z_\mu$,$W_\mu$
is the polarization 4-vector of the $Z^0$- and $W^{\pm}_L$-
bosons, $ a_{L}=\frac{-\frac{1}{2}+\sin^2\theta_W }{\sin \theta _W\cos\theta _W},
a_{R}=\tan\theta_W$,
 $s=(k_1+k_2)^2$,
 $t=(k_1-k_4)^2$,
 $u=(k_2-k_4)^2$,
 $m_H$
is the  mass of $\delta^{--}_{L,R}$-bosons,
$\Gamma_H$ is the width of the decay $\delta^{--}_L\rightarrow \delta^-_LW^-_L$ .

For the differential cross section we obtain 
the following result:

\begin{equation}
\label{A8}
\frac{d\sigma }{dt}=\frac{1}{2}\frac{\alpha h_{ee}^2}{s}^2\,
\frac{1}{4\sin^2\theta_W}\left(
a\left(\frac{1}{t}+\frac{1}{u}\right)+\left(2b
-4\right)-m_W^2m_h^2\left(
\frac{1}{t^2}+\frac{1}{u^2}\right)\right),
\end{equation}
\begin{equation}
\label{A9}
a=m_W^2+m_h^2-s-\frac{4sm^2_h(s-m_H^2)}{(s-m_H^2)^2+m_H^2\Gamma_H^2}-\frac{2(s-m^2_W)m_H^2}
{m_W^2+m^2_h-s},
\end{equation}
\begin{equation}
\label{A10}
b=\frac{s}{m_W^2}\left(\frac{\left(\left(m_H^2
-m_h^2-m_W^2\right)^2-4m^2_{h}m^2_{W}\right)+m_H^2\Gamma_H^2}
{\left(s-m_H^2\right)^2+m^2_H\Gamma^2_H}\right),
\end{equation}
\begin{equation}
\label{A11}
t_-<t<t_+,
\end{equation}
\begin{equation}
\label{A12}
t_{\pm}=\frac{m_h^2+m_W^2-s\pm
\sqrt{(m_h^2+m_W^2-s)^2-4m_h^2m_W^2}}{2}.
\end{equation}

Here $m_h$ denotes the mass of the singly triplet charged Higgs
bosons.

Integrating within the limits (13),(14) we obtain for the total
cross section the following result:
\begin{equation}
\label{A13}
\sigma =\frac{\alpha h_{ee}^2}{s}\,\frac{1}{
2\sin^2\theta_W}(a\log (\frac{t_+^2}{m_h^2 m_W^2})+(b-3)(t_+-t_-)).
\end{equation}
At $\sqrt{s} \gg m_H,m_h,m_W $ the previous
 formula is reduced and we have:
\begin{equation}
\label{A14}
\sigma =\frac{\alpha h_{ee}^2}{s}
\frac{1}{2\sin^2\theta_W}
(2\log(\frac{s}{m_h m_W})-3).
\end{equation}
The differential and the total cross section of the processes (3) may be
obtained from the
formulas (10)-(16) at $m_H=m_h$, $\Gamma_H=0$ by the following replacements:

\begin{equation}
\label{A15}
\frac{1}{2\sin^2\theta_W}\rightarrow a_{L,R}^2, m_W\rightarrow
m_Z.
\end{equation}

On Fig. 2,3 we present
the number of events
$\delta^{--}_{L}Z^0$ and
$\delta^{-}_{L}W^-_L $  
per year for the processes (3) and (4) versus $m_H$ at fixed
$\sqrt{s}$ and fixed $m_h$ at yearly luminosity  $L=100 fb^{-1}$
($e^-e^-$-colliders with yearly
luminocity  $L=100 fb^{-1}$ has been considered e.g. in
ref.\cite{S}).

The number of events
$\delta^{--}_{R}Z^0$ may be easily obtained from Fig.2 by multiplying
the number of events
$\delta^{--}_{L}Z^0$ by $\frac{a^2_R}{a^2_L}=0.742$.  On the other hand at the
$m_H=m_h$ 
$\delta^{-}_{L}$- 
 bosons may be produced more efficiently (by $\frac{1}{4\sin^2\theta_W
a_{L,R}^2}=2.661,3.586$ times)  in
 reaction (4) than $\delta^{--}_{L,R}$-bosons
in reaction (3).Indeed, at 
$m_H=m_h$ 
the cross sections of processes (3) and (4) are different from each other
only by coefficients and the influence of the $m_W$ and $m_Z$ mass difference
far from the threshold is negligible.

{\bf 3.Comparision with other mechanisms of triplet Higgs bosons production}

The processes
\begin{equation}
\label{A16}
  e^+e^-\rightarrow
\delta^{++}_{L,R}\delta^{--}_{L,R},\delta^{+}_{L}\delta^{-}_{L}.
\end{equation}

have  a large cross
 section \cite{TR}-\cite{S1}\cite{FC}-\cite{JG},\cite{S2},\cite{S3}, however, it becomes
 kinematically allowed at energies
  $\sqrt{s}>2m_H$,
whereas process (3),(4) is kinematically allowed at lower
energies   $\sqrt{s}>m_H+m_Z,m_h+m_W$.

From formula (2.1) of the ref.\cite{G}  and from Fig.2 we see that at
$\sqrt{s}=1$ TeV, far from the threshold  the number 
$\delta^{++}_{L,R}\delta^{--}_{L,R}$ pairs  produced in reaction (18) 
exceeds by approximately 1500 times the number of 
$\delta^{--}_{L}$ bosons produced in reaction (3) and the number of 
$\delta^{+}_{L}\delta^{-}_{L}$ pairs produced in reaction (19)
 as seen from Fig.14 of the ref.\cite{FC} and from Fig.3
exceeds by approximately 100 times the number of 
$\delta^{-}_{L}$ bosons produced in reaction (4) at $\sqrt{s}=1$ TeV.

It must be noted, however, that  with the growth
of $h_{ee}$  the number of the  triplet Higgs bosons
 produced in
 reactions (3),(4) quadratically increases.

Besides, the resonant enhancement (if $m_H>m_h+m_W$) in the process (4), 
as seen from Fig.3, may considerably increase the cross section of the
 process (4)
in comparision with the process (19) even sufficiently far from resonance.

It must also be noted, that the processes (18),(19) decrease with the growth of $\sqrt{s}$
as 
$s^{-1}$
 whereas, as seen from formula (16) the processes (3,4) decrease more slowly
as $s^{-1}\log(\frac{s}{m_h m_W})$.

At LHC  doubly charged Higgs bosons may be produced in
pairs  \cite{G, JG}; at
integrated  luminosity  $L=100fb^{-1}$, $\sqrt{s}=14$ TeV,
and $m_H=800$ GeV we
 have about 100 $\delta^{++}_{L}\delta^{--}_{L}$ 
 pairs per  year
 which is comparable with the number of
$ \delta^{--}_{L}$-bosons produced in reaction (3) at
 $\sqrt{s}=1$ TeV, $L=100fb^{-1}$, $h_{ee}=10^{-2}$ and the same mass of the
 doubly charged Higgs bosons.
 The number of $\delta^{+}_{L}\delta^{-}_{L}$
pairs produced at  LHC several times is smaller than
number of $\delta^{++}_{L}\delta^{--}_{L}$ bosons whereas about
200
of $ \delta^{-}_{L}$-bosons may be produced in reaction (4)
at  $\sqrt{s}=1$ TeV, $L=100fb^{-1}$, $h_{ee}=10^{-2}$,
 and the same mass of the
 triplet charged Higgs bosons if resonance is absent.

Triplet Higgs bosons may be produced also in $e^-\gamma$-
collisions in reaction
\begin{equation}
\label{A18}
e^-\gamma
\rightarrow \delta^{--}_{L,R}l^+,
 \end{equation}

for doubly charged Higgs bosons case
\cite{FC},\cite{LTND},\cite{BHMR} and in
the reaction

\begin{equation}
\label{A19}
e^-\gamma
\rightarrow \delta^{-}_L\nu,
 \end{equation}

for singly charged Higgs bosons case \cite{FC}.

As seen from Fig.2 of the ref.\cite{LTND},
at $L=100 fb^{-1}$,
$h_{ee}=10^{-2}$
 about
700-350, 200-90 (at $\sqrt{s}=0.5,1$ TeV respectively) 
doubly charged Higgs bosons with 0.1 TeV $<m_H$
may be produced per year in reaction (20).
The number of $ \delta^{-}_{L}$-bosons produced in
reaction (21) at  the same energies, luminosities (and far from
resonance) is of the same
order.

{\bf 4.Acknowledgements}

The author express his sincere gratitude to I.G.Aznauryan 
for helpful discussions
 and to F. Cuypers  for helpful discussions and for kind
 hospitality at Paul Scherrer Institute.

\setcounter{equation}{0}
\appendix{{\bf Appendix A}}
\renewcommand{\theequation}{A.\arabic{equation}}
\indent
In the left-right symmetric
model with Higgs triplets the matrix
 of left and right Higgs triplets with
($T=1,Y=2$) may be written as:
\begin{equation}
\label{AD}
\Delta  _{L,R}= \left(\begin{array}{ll}
 \delta  ^+_{L,R}/\sqrt{2}&
\delta  ^{++}_{L,R}\\
\delta  ^0_{L,R}& - \delta^+_{L,R}/\sqrt{2}
\end{array}\right)
\end{equation}
and their interaction with the left- and right-handed lepton fields $\psi^T_{L,R}=(\nu
_{L,R}^T,e^T_{L,R}) $ is  described by the
lagrangian:
\begin{equation}
\label{DD}
{\cal L}=  ih _{ij}\left(\psi^ T_{iL}
C \tau _2\Delta _L\psi_{jL}
+\psi^{T}_{iR} C\tau _2 \Delta _R \psi_{jR} \right)+ h.c.
\end{equation}

Here $i$,$j=e,\mu,\tau $ are generations indices, $C$ is the  charge
conjugation matrix, and $\tau_2$ is the Pauli matrix.
After symmetry breaking Majorana masses of the heavy approximately
right-handed neutrinos are expressed through the Yucawa couplings
$h$ and the neutral component of the right triplet vacuum expectation value
($v_R$)
in the following  way:

\begin{equation}
\label{DE}
m_{N}=\sqrt{2}hv_{R}.
\end{equation}
Also, large right triplet vacuum expectation (
$v_L\ll k_L$,$k_R\ll v_R$,
$k_L,k_R$- are vacuum expectations of the left and right doublets,
$v_L$-vacuum expectation of the left triplet) 
 provides the mass of the
$W_{R}^{\pm}$-bosons:
\begin{equation}
\label{DF}
m_{W_{R}}=\frac{1}{2}gv_R
\end{equation}
whereas the doublet vacuum expectation is responsible for the mass of
the $W_{L}^{\pm}$-bosons.

From (A2) we derive charged triplet Higgs bosons interactions
 with leptons:

\begin{equation}
\label{DK}
{\cal{L}}=
-\sqrt{2}h_{ee}\bar{l^c}P_{L}\nu\delta^{+}_L+h.c.-
h_{ee}\bar{l^c}P_{L,R}l\delta^{++}_{L,R}
\end{equation}
where $l^c=C\bar{l}^T$.

Vertexes $\gamma \delta^{++}_{L,R} \delta^{--}_{L,R}$,$Z^0
\delta^{++}_{L,R} \delta^{--}_{L,R}$,$W^-_L\delta^{-}_{L}
\delta^{++}_{L}$ are described
 by the following Feynmann rules:
$i2e\left(p-q\right)^{\mu}\cdot A_{\mu}$, 
 $i2a_{L,R}e\left(p-q\right)^{\mu}\cdot Z_{\mu}$, 
 $ie\frac{1}{\sin\theta_W}
\left(p-q\right)^{\mu}\cdot W_{\mu}$ where $p,q$- denotes
 incoming momentums of the Higgs bosons.

 Doubly charged Higgs bosons
 mass matrix 
is nondiagonal 
, however, in a wide range of
Yucava coupling and vacuum expectations 
however, in more general case left and right
doubly charged Higgs bosons 
may be, in principle, mixed.

Mass matrix of the doubly and singly charged Higgs bosons in general is nondiagonal
 (see formula (A4),(A5) in ref.\cite{G}), however in
the left-right symmetric model
$\delta^{++}_{L}$ - $ \delta^{--}_{R}$ and doublet-triplet mixing
 is negligible because it is proportional to the $v_L$.

In the above mentioned models where $v_L$ is not restriced by experiment there may a large
mixing among  the doublet and triplet singly charged Higgs bosons .

 In the end  we must remind that in the limit $v_L\ll k_L$,$k_R\ll v_R$
$ \delta^{-}_{R}$-boson is approximately the charged Goldstown boson and
after spontaneous symmetry breaking it becomes the longitudional part of the
$W_R^{\pm}$- bosons.

\centerline {\bf Figures captions:}

Fig.1 Diagrams corresponding to the processes (2)-(4).

Fig.2 Number of events 
$\delta^{--}_{L,R}Z^0 $ per year ($\sigma L$) 
versus $m_H$ at fixed $\sqrt{s}$
(at yearly luminosity  $L=100 fb^{-1}$) produced in reaction (3)
as a function of $m_{H}$ with $h_{ee}=10^{-2}$. Curves 1,2
correspond to the 
$\sqrt{s}= 0.5, 1$  TeV 
respectively.

Fig.3 Number of events
$\delta^{-}_{L}W^-_L $ 
per year
 ($\sigma L$) versus $m_H$ at fixed $\sqrt{s}$ produced in
reaction (4)
(at yearly luminosity  $L=100 fb^{-1}$).
as a function of $m_{H}$ with $h_{ee}=10^{-2}$.Curves 1,2
correspond to the number of events at  $m_H=1.3m_h$
and $\sqrt{s}= 0.5, 1$  TeV 
 respectively.
Curves 3,4 correspond to the
number of events $\delta^{-}_{L}W^-_L $ events per year at
 $m_H=m_h$
and energies $\sqrt{s}=0.5,1$ TeV respectively.
\newpage
\setcounter{figure}{0}
\begin{figure}
\begin{center}
\epsfxsize=10.cm
\leavevmode\epsfbox{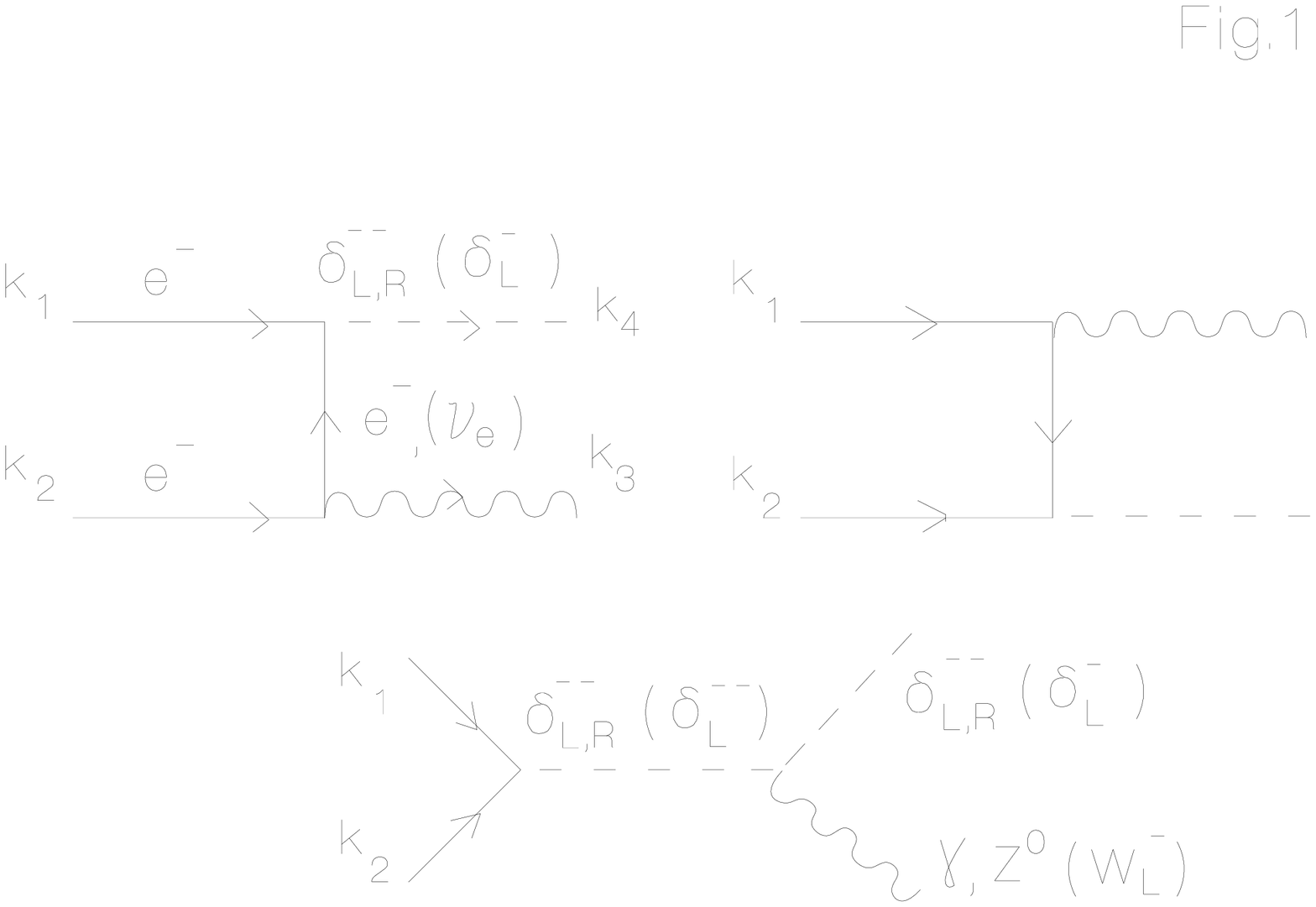}
\end{center}
\caption{}
\end{figure}

\begin{figure}
\begin{center}
\epsfxsize=10.cm
\leavevmode\epsfbox{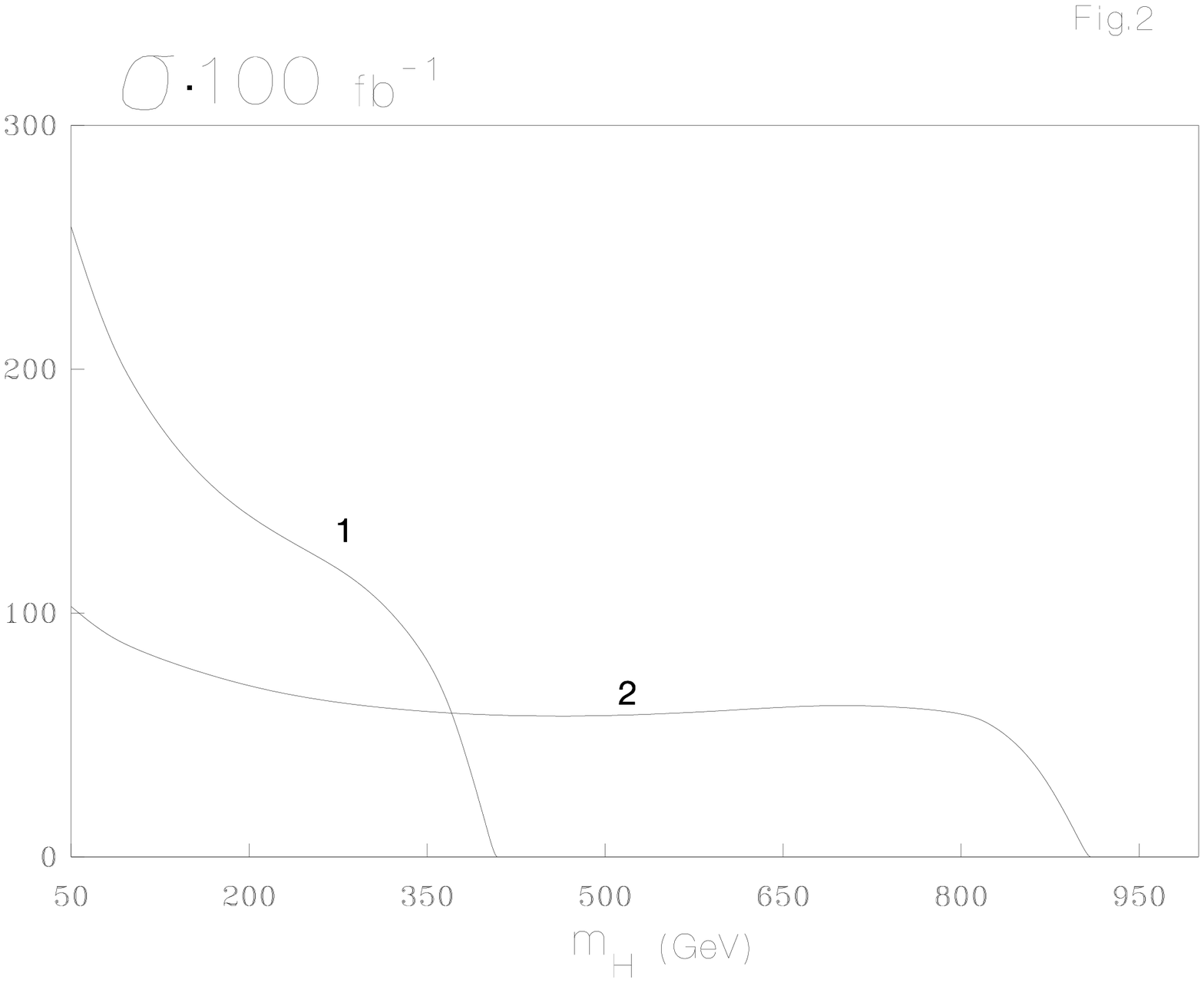}
\end{center}
\caption{}
\end{figure}

\newpage

\begin{figure}
\begin{center}
\epsfxsize=10.cm
\leavevmode\epsfbox{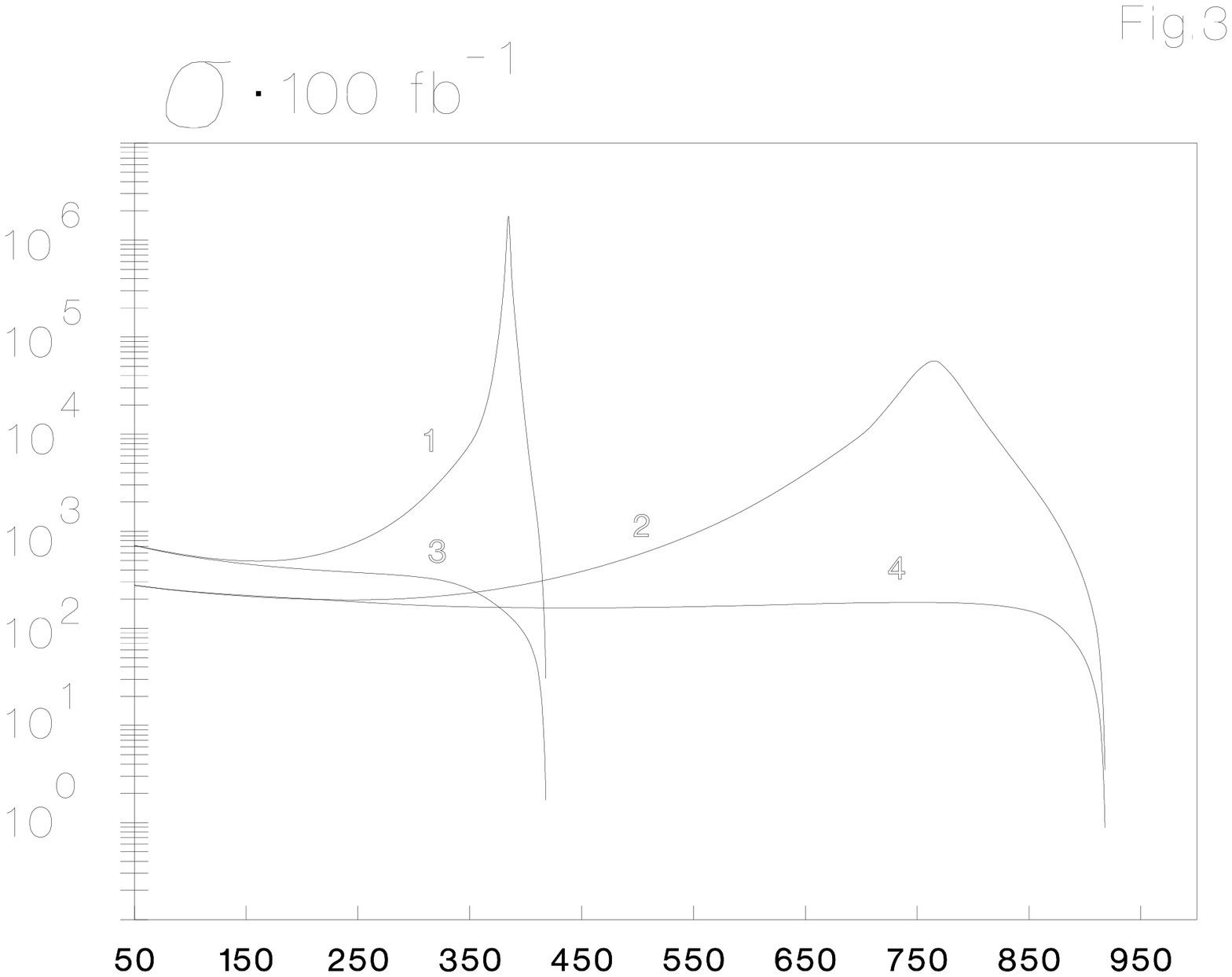}
\end{center}
\caption{}
\end{figure}

\end{document}